\documentclass[twocolumn,prl,superscriptaddress,showpacs,preprintnumbers,amsmath,amssymb]{revtex4}
\usepackage[ansinew,latin1]{inputenc}
\usepackage{psfrag}
\usepackage{array}
\usepackage{amssymb}
\usepackage{amsmath}

\usepackage{graphicx}

\begin{document}

\title{Test of Lorentz invariance with spin precession of ultracold neutrons}

\def\ECTUM{Excellence Cluster `Universe', \TUM}
\def\GUM{Johannes--Gutenberg--Universit\"at, D--55128 Mainz, Germany}
\def\HNINP{Henryk Niedwodnicza\'nski Institute for Nuclear Physics, 31--342 Cracow, Poland}
\def\ILL{Institut Laue--Langevin, F--38042 Grenoble Cedex, France}
\def\JENA{Department of Neurology, Friedrich--Schiller--University, Jena, Germany}
\def\JINR{JINR, 141980 Dubna, Moscow region, Russia}
\def\JUC{Marian Smoluchowski Institute of Physics, Jagiellonian University, 30--059 Cracow, Poland}
\def\KCGUM{Institut f\"ur Kernchemie, \GUM}
\def\KULEUVEN{Instituut voor Kern-- en Stralingsfysica, Katholieke~Universiteit~Leuven, B--3001 Leuven, Belgium}
\def\LPC{LPC Caen, ENSICAEN, Universit\'e de Caen, CNRS/IN2P3, F--14050 Caen, France}
\def\LPSC{LPSC, Universit\'e Joseph Fourier Grenoble 1, CNRS/IN2P3, Institut National Polytechnique de Grenoble 53, F--38026 Grenoble Cedex, France}
\def\PGUM{Institut f\"ur Physik, \GUM}
\def\PSI{Paul Scherrer Institut (PSI), CH--5232 Villigen PSI, Switzerland}
\def\RAL{Rutherford Appleton Laboratory, Chilton, Didcot, Oxon OX11 0QX, United Kingdom}
\def\SUSSEX{Department of Physics and Astronomy, University of Sussex, Falmer, Brighton BN1 9QH, United Kingdom}
\def\TUM{Technische Universit\"at M\"unchen, D--85748 Garching, Germany}
\def\UNIFR{University of Fribourg, CH--1700, Fribourg, Switzerland}

\author{I. Altarev}      	\affiliation{\TUM}
\author{C. A. Baker}     	\affiliation{\RAL}
\author{G. Ban}          	\affiliation{\LPC}
\author{G. Bison}        	\affiliation{\JENA}
\author{K. Bodek}        	\affiliation{\JUC}
\author{M. Daum}         	\affiliation{\PSI}
\author{P. Fierlinger}   	\affiliation{\ECTUM}
\author{P. Geltenbort}   	\affiliation{\ILL}
\author{K. Green}        	\affiliation{\RAL} \affiliation{\SUSSEX}
\author{M. G. D. van der Grinten} \affiliation{\RAL} \affiliation{\SUSSEX}
\author{E. Gutsmiedl}    	\affiliation{\TUM}
\author{P. G. Harris}    	\affiliation{\SUSSEX}
\author{W. Heil}         	\affiliation{\PGUM}
\author{R. Henneck}      	\affiliation{\PSI}
\author{M. Horras}       	\affiliation{\ECTUM} \affiliation{\PSI}
\author{P. Iaydjiev}     	\altaffiliation{On leave of absence from INRNE, Sofia, Bulgaria} \affiliation{\RAL} 
\author{S. N. Ivanov}    	\altaffiliation{On leave from PNPI, Russia} \affiliation{\RAL}
\author{N. Khomutov}     	\affiliation{\JINR}
\author{K. Kirch}        	\affiliation{\PSI}
\author{S. Kistryn}      	\affiliation{\JUC}
\author{A. Knecht}       	\altaffiliation{Also at University of Z\"urich} \affiliation{\PSI}
\author{P. Knowles}      	\affiliation{\UNIFR}
\author{A. Kozela}       	\affiliation{\HNINP}
\author{F. Kuchler}     	\affiliation{\ECTUM}
\author{M. Ku\'zniak}    	\affiliation{\JUC} \affiliation{\PSI}
\author{T. Lauer}        	\affiliation{\KCGUM}
\author{B. Lauss}        	\affiliation{\PSI}
\author{T. Lefort}       	\affiliation{\LPC}
\author{A. Mtchedlishvili}  \affiliation{\PSI}
\author{O. Naviliat-Cuncic} 	\affiliation{\LPC}
\author{A. Pazgalev}     	\affiliation{\UNIFR}
\author{J. M. Pendlebury}	\affiliation{\SUSSEX}
\author{G. Petzoldt}     	\affiliation{\PSI}
\author{E. Pierre}       	\affiliation{\LPC} \affiliation{\PSI}
\author{G. Pignol}       	\altaffiliation{pignol@lpsc.in2p3.fr} \affiliation{\LPSC} 
\author{G. Qu\'em\'ener} 	\affiliation{\LPC} 
\author{M. Rebetez}      	\affiliation{\UNIFR} 
\author{D. Rebreyend}    	\altaffiliation{rebreyend@lpsc.in2p3.fr} \affiliation{\LPSC}
\author{S. Roccia}       	\altaffiliation{roccia@lpsc.in2p3.fr} \affiliation{\LPSC}
\author{G. Rogel}        	\affiliation{\LPC} \affiliation{\ILL}
\author{N. Severijns}    	\affiliation{\KULEUVEN}
\author{D. Shiers}       	\affiliation{\SUSSEX}
\author{Yu. Sobolev}     	\affiliation{\PGUM}
\author{A. Weis}         	\affiliation{\UNIFR}
\author{J. Zejma}        	\affiliation{\JUC}
\author{G. Zsigmond}     	\affiliation{\PSI}

\date{\today}

\begin{abstract}
A clock comparison experiment, analyzing the ratio of spin precession frequencies 
of stored ultracold neutrons and $^{199}$Hg atoms is reported. 
No daily variation of this ratio could be found, from which is set an upper limit 
on the Lorentz invariance violating cosmic anisotropy field 
$b_{\bot} < 2 \times 10^{-20} \ {\rm eV}$ (95\% C.L.). 
This is the first limit for the free neutron. 
This result is also interpreted as a direct limit on the gravitational 
dipole moment of the neutron 
$|g_n| < 0.3 \ $eV/$c^2$\,m 
from a 
spin-dependent interaction with the Sun. 
Analyzing the gravitational interaction with the Earth, based on previous data,
yields a more stringent limit 
$|g_n| < 3 \times 10^{-4} \ $eV/$c^2 \ $m. 
\end{abstract}


\maketitle

Lorentz symmetry is a fundamental hypothesis of our current
understanding of physics and is central to the foundations
of the Standard Model of particle physics (SM).
However, the SM is widely believed to be only the low energy limit of
some more fundamental theory, a theory which
will probably violate more symmetries than the SM, in order to
accomodate some features of the universe currently lacking in the SM,
e.g., the baryon asymmetry.
A SM extension including Lorentz and CPT violating terms has been
presented in~\cite{Colladay}.
It provides a parametrisation of effects suitable to be tested by low energy precision experiments.
In particular, clock comparison
experiments~\cite{Lamoreaux1995,Bear2000} have proven to be
particularly sensitive to spin-dependent effects arising from a
so-called \emph{cosmic spin anisotropy field} $\tilde{\bf b}$ filling the whole universe. 
This Letter reports on a search for such an exotic field via its
coupling to free neutrons.

In the presence of a 
field $\tilde{\bf b}$, the two spin states of the neutron will 
encounter an extra contribution to the
energy splitting corresponding to the potential 
$V = {\bf \sigma}~\cdot~\tilde{\bf b}$ where $\sigma$ are the Pauli matrices. 
Thus, if a neutron is subjected to both a static magnetic field ${\bf B}$ and the new field $\tilde{\bf b}$, 
its spin will precess at the modified Larmor frequency $f_n$, 
which to first order in $\tilde{b}$ is given by
\begin{equation}
f_n = \frac{\gamma_n}{2 \pi} B + \frac{2}{h} \tilde{\bf b} \cdot \frac{\bf B}{B}. 
\end{equation}

We searched for a sidereal modulation (at a period of $23.934$~hours) 
of the neutron Larmor frequency induced by  $b_{\bot}$, 
the component of $\tilde{\bf b}$ orthogonal to the Earth's rotation axis. 
The experiment is also sensitive to
a possible influence of the Sun 
on the spin precession dynamics, leading to a solar modulation (at a period of 24~h) 
of the Larmor frequency, as proposed in \cite{Gol78}. 
Such an effect could arise from a non-standard spin-dependent component 
of gravity~\cite{Leitner:1964,HariDass} or from another long-range spin-dependent 
force~\cite{Dobrescu:2006au,Moody:1984ba}. 
In particular, a non-zero \emph{neutron gravitational dipole moment} $g_n$ would induce 
a coupling through (see also~\cite{Gol91})
\begin{equation}
V_{\rm gdm} 
= g_{\rm n} \frac{G M}{r^3} \ {\bf \sigma} \cdot {\bf r}, 
\label{Eq0}
\end{equation}
where $G$ is Newton's constant, 
and for the mass $M$ and the distance $r$ we use
the Sun mass $M_{\odot}$ 
and the distance Earth--Sun $r_{\odot}$.

The experimental apparatus at the PF2~\cite{PF2} beamline at ILL,
Grenoble, is normally used to search for the electric dipole moment of
the neutron (nEDM)~\cite{Harris, Baker2006}.
The apparatus permits spin-polarized ultracold neutrons (UCN) to be
filled into a volume, stored, and then
counted and classified by polarization state.
While confined, the neutrons can be exposed to static (normally
$B\approx 1~\mu$T), as well as to oscillating, magnetic fields.
A surrounding four layer Mu-metal shield suppresses the external
magnetic field and its fluctuations.
Although $\tilde{\bf b}$ acts like a magnetic field influencing the particles' spin precession, 
it can, per definition, not be suppressed by the Mu-metal shield. 

\begin{figure}
\includegraphics[width=0.77\linewidth]{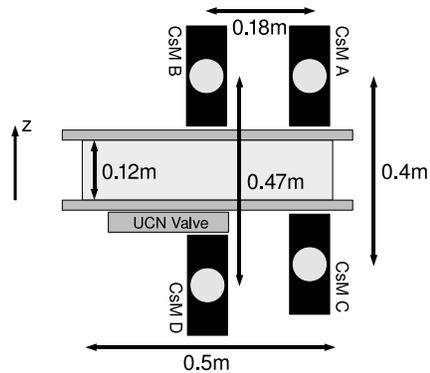}
\caption{
Vertical cut through the cylindrical precession chamber for
  UCN and $^{199}$Hg.  
Schematically indicated are the
  $\approx70$~mm diameter Cs vapor filled bulbs and their mounts.  The
  scalar Cs magnetometer measures the magnitude of $B$ found
  at the center of the spherical bulb.
} \label{chamberandcesiums}
\end{figure}

The neutron Larmor frequency, $f_{\rm n} = \gamma_{\rm n} B /(2 \pi)
\approx 30$\,Hz, is measured via the Ramsey method of separated
oscillatory fields~\cite{K. Green,Ramsey}.
Following filling, an initial oscillating field pulse
rotates the neutron spin by $\pi/2$, leaving the magnetic moment at
right angles to the static holding field $B$, whereupon it precesses.
Following a free spin precession time of typically 100~s, a second oscillating field pulse,
phase coherent with the first pulse, further rotates the neutron spin by $\pi/2$. 
%
%
The accumulated phase is measured by counting the populations of
the two resulting spin states following the second Ramsey pulse.
For each cycle about $10^4$
neutrons are counted allowing a measurement of $f_{\rm n}$ with a
statistical precision of $\Delta f_{\rm n} \approx 50~\mu$Hz.
A unique feature of this nEDM apparatus is the use of a mercury
co-magnetometer~\cite{K. Green}.
Within the neutron precession chamber, nuclear spin polarized
$^{199}$Hg atoms precess in the same $B$ field as the neutrons.
The Larmor frequency $f_{\rm Hg} = \gamma_{\rm Hg} B /(2 \pi) \approx
8$\,Hz is measured optically  for each cycle to a precision of $\Delta
f_{\rm Hg} \approx 1~\mu$Hz.
The pumping and analysing light are generated by two lamps filled with $^{204}$Hg and Ar plasma. 
In addition, four scalar Cs magnetometers \cite{Groeger} are placed
above and below the precession chamber (see Fig.~\ref{chamberandcesiums}).
They provide on-line measurements of the magnetic field with a
precision of 150~fT and were used to measure the vertical gradients
of the magnetic field.

For the clock comparison experiment, we use 
the ratio $R = f_{\rm n}/f_{\rm Hg}$ which 
suppresses the effect of magnetic field fluctuations in the limit of a perfectly homogeneous field. 
The existing constraints for $^{199}$Hg~\cite{Lamoreaux1995} are sufficiently tight, so within
the sensitivity of this experiment, new physics effects can only show up in $f_{\rm n}$.
While the Earth is rotating, together with the vertical quantization axis, the 
new physics effects under consideration would appear in a harmonic change of $R$:
\begin{equation}
\label{Eq1}
R(t) = \left | \frac{\gamma_{\rm n}}{\gamma_{\rm Hg}} \right | + A \sin \left( 2 \pi t / T + \phi \right) + \delta R.
\end{equation}

The constant term $\delta R$ would be induced by the component of a new field parallel to the Earth rotation axis, 
while the amplitude $A$ would be induced by the orthogonal component. 
Since magnetic field gradients related effects can be mistaken for the steady term $\delta R$, 
we first 
focus on the search for a non zero amplitude $A$. 

\begin{figure}
\includegraphics[width=1.\linewidth]{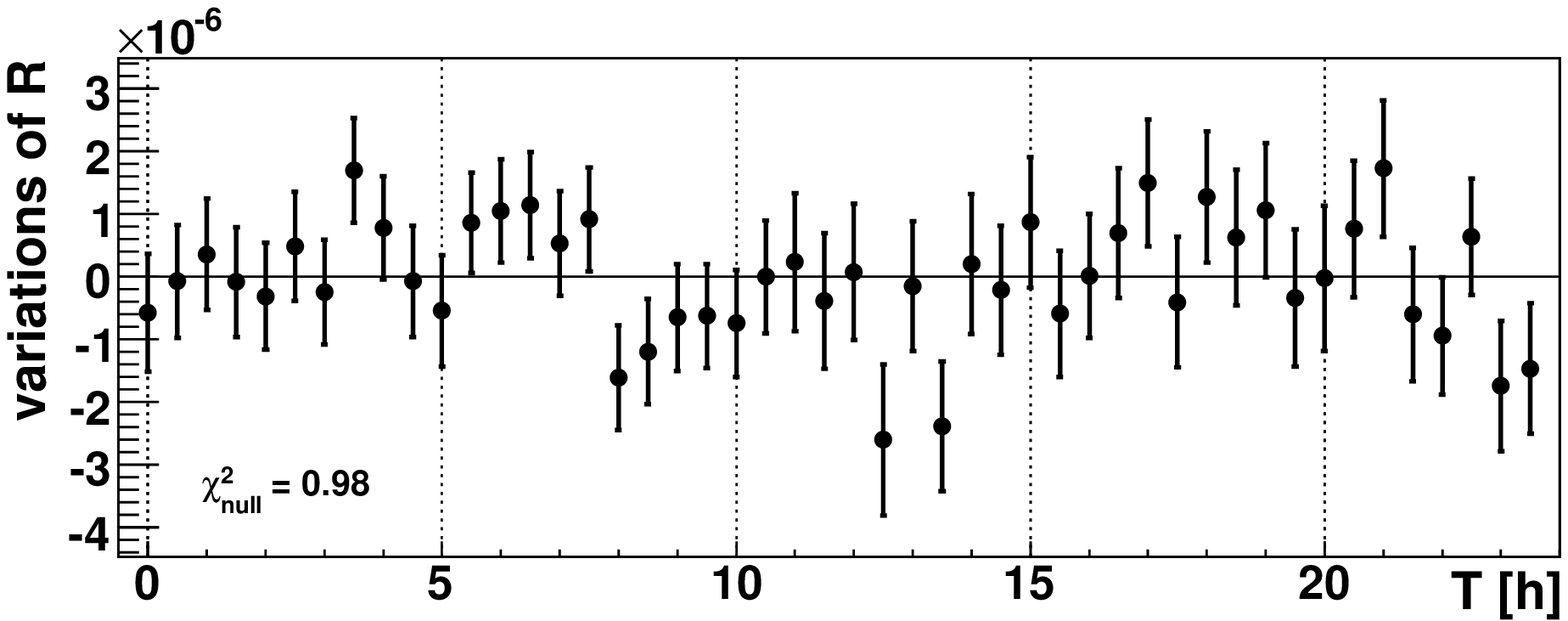}
\includegraphics[width=1.\linewidth]{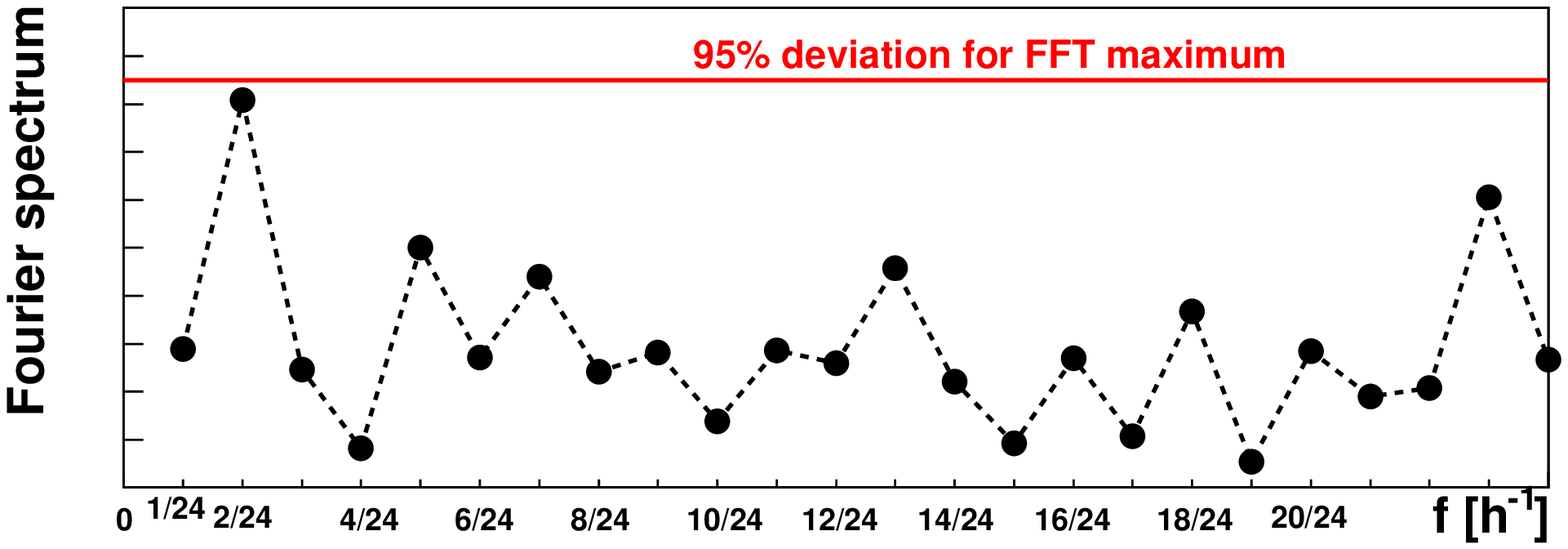}
\caption{The upper figure shows the variation of the ratio $R$ around its average. 
For clarity, the data are folded modulo 24 hours and binned every half hour.
The lower figure shows modulus of the discrete Fourier Transform of the same dataset. 
  the line would be an evidence that this frequency is too much represented as compared with a white noise. } \label{Data}
\end{figure}

Data were recorded in April--May 2008 
with the $B$ field pointing downwards.
The first 35~h of data were recorded starting on April 21 07h20 UT, followed by a
break of 255~h, and then 85~h of uninterrupted data were
collected.
As we search for a signal modulation in $R(t)$, the runs were combined after subtracting the mean values 
$\bar{R}$ of the corresponding runs. 
Figure \ref{Data} shows folded data, and its discrete spectral analysis. 
The error bars indicate combined statistical errors of the neutron and the Hg frequency extraction, 
dominated by the former one. 
The spectral analysis shows that no particular frequency 
can be extracted from the data and the whole dataset is compatible with a signal of null amplitude  
($\chi^{2}_{\mbox{\scriptsize{null}}}=0.98$).

The neutron frequency extraction \cite{K. Green} requires a fit of the visibility $\alpha$ of the
Ramsey resonance curve. The value of this parameter depends on the value of the magnetic field gradients. 
In order to avoid systematic effects correlated with the value of these gradients
(which could be daily modulated), $\alpha$ was fitted in small (typically one hour) subsets of data 
 to ensure that the 
neutron frequency extraction does not create any bias. 

The $R$ parameter also depends on the value of the magnetic field gradients inside the chamber: 
while the center of mass of the thermal $^{199}$Hg gas coincides with the chamber center, the UCN
center of mass is about 
$h=2.8$\,mm lower, due to gravity. 
This offset is related to the vertical 
gradient $\partial B / \partial z$~\cite{Baker2006}:
\begin{equation}
R = \left | \frac{\gamma_{\rm n}}{\gamma_{\rm Hg}} \right | \left( 1 + \frac{(\partial B / \partial z) h }{B} \right).
\label{RCurve}
\end{equation}
Daily variations of  $\partial B / \partial z$ would be the main systematic uncertainty in our analysis and  
could appear, e.g., due to a daily modulation of the Earth magnetic field. 
Therefore, the vertical gradients were monitored by the Cs magnetometers 
(two on top of the storage chamber and two below it). 
At the frequency of interest, 1/24~h, the amplitude of fluctuations of $\partial B / \partial z$
was mesured to be $\leq 20$\,pT/m 
resulting in a daily modulation of $R$ with an amplitude $\leq 2 \times 10^{-7}$, according to Eq. (\ref{RCurve}). 
This effect is small enough to prevent the $R$ ratio departing from a white noise signal, as one can see in Fig. \ref{Data}. 
Other possible sources of false daily modulated signal 
have been investigated and ruled out. 
Besides magnetic field inhomogeneities, the main remaining effect is related to the light shift of the mercury frequency. 
We estimated the associated relative error to be $\approx 10^{-7}$ with our analyzing light intensity. 
The drifts in intensity of the light was measured to be less than 10~\%.

To extract an upper limit for the daily modulation amplitude, a frequentist confidence level analysis was performed on the unfolded data. 
The method consists in determining wether a given signal hypothesis 
(a given amplitude $A$ and phase $\phi$) can be excluded at $95 \%$ C.L. when compared with the null hypothesis.
This method is known to optimally discriminate two signal hypotheses \cite{PDG}.
For a given $A$ and $\phi$, we form the quantity:
\begin{eqnarray}
Q(A, \phi) & = & \chi^2_{\rm null} - \chi^2_{\rm signal} \\
\nonumber
\chi^2_{\rm null}   & = & \frac{1}{N} \sum_{i = 1}^N \left( \frac{R[i]}{\Delta R[i]} \right)^2 \\
\nonumber
\chi^2_{\rm signal} & = & \frac{1}{N} \sum_{i = 1}^N \left( \frac{R[i] - A \sin(2 \pi t[i] / T + \phi)}{\Delta R[i]} \right)^2
\end{eqnarray}
where $N = 2070$ is the total number of data points, $R[i]$ is the R ratio subtracted 
from its mean value in individual datasets and $T$ is either the solar period or the sidereal
period. 
From the measured data, $Q_{\rm data}(A, \phi)$ is calculated. 
We consider the probability distribution of $Q$, $\rho_{\rm null}(Q)$ in the null hypothesis and $\rho_{\rm signal}(Q)$ in the signal hypothesis. 
These two probability distributions have been calculated (for each signal hypothesis) using Monte-Carlo simulations. 
The confidence level of the hypothesis $(A, \phi)$ is then defined as
\begin{equation}
CL(A, \phi) = \int_{-\infty}^{Q_{\rm data}} \rho_{\rm signal}(Q) dQ.
\end{equation}
The amplitude $A$ is excluded at $95 \%$~C.L. if, for all phases $\phi$, we have $CL(A, \phi) < 0.05$.
The statistical limit obtained this way is $A \  <  0.58 \times 10^{-6}$, both, for the sidereal and solar period. 
Accounting for the systematics that $\partial B / \partial z$ modulations could counteract any other modulation, the amplitude due to new physics is limited to: 
\begin{equation}
\label{resultA}
A        \  <  0.8 \times 10^{-6} \quad  95 \% \ {\rm C.L.} 
\end{equation}

Our result can be interpreted in terms of a limit on the cosmic spin anisotropy field for the free neutron.
In this case, $T$ is the sidereal period and the amplitude $A$ is related to the $b_{\bot}$ component according to: 
\begin{equation}
\label{A_b}
A = 2 b_{\bot} \frac{\cos(\lambda)}{h \, f_{\rm Hg}}
\end{equation}
where $\lambda = 45^{\circ} 12'22''$ is the latitude of the experiment in Grenoble and $h$ is the Planck constant;
thus:
\begin{equation}
b_{\bot} \  < 2 \times 10^{-20} \ {\rm eV}  \quad  95 \% \ {\rm C.L.}
\end{equation}
Table \ref{ALLResults} compares this result to existing limits on other particles. 
The result reported here is the first limit for the free neutron. 
It is complementary 
to the more precise atomic experiments~\cite{Lamoreaux1995,Bear2000} 
that can be interpreted as limits concerning bound neutrons inside nuclei. 
Contrary to the results of \cite{Lamoreaux1995, Bear2000}, 
the neutron result is free from model-dependent nuclear corrections and possible related suppression effects.  
Being a factor 200 more stringent than the limit for the proton, it is the best free nucleon limit to date.

\begin{table}[hhh]
\begin{center}
\begin{tabular}{|c|c|c|c|}
\hline
Reference	 				&System 	& Particle 		 & $b_{\bot}$ [eV]     \\
\hline 
Berglund \textit{et al.}, \cite{Lamoreaux1995}	& Hg $\&$ Cs 	& bound neutron          & $9 \times 10^{-22}$ \\
						& 	 	& electron               & $2 \times 10^{-20}$ \\
Bear \textit{et al.}, \cite{Bear2000}		& Xe $\&$ He	& bound neutron          & $2 \times 10^{-22}$ \\
Phillips \textit{et al.},\cite{Phillips}	& H		& proton		 & $4 \times 10^{-18}$ \\
Heckel \textit{et al.},	\cite{Heckel:2006ww}    & e		& electron               & $7 \times 10^{-22}$ \\
Bennet \textit{et al.},  \cite{muon}         	& $\mu$         & positive muon          & $2 \times 10^{-15}$ \\
                                                &               & negative muon          & $3 \times 10^{-15}$ \\
{\bf This analysis}				& n $\&$ Hg	& free neutron	 	 & $2 \times 10^{-20}$ \\ 
\hline
\end{tabular}
\caption{\label{ALLResults} Results of more restricting upper limits (at 95$\%$~C.L.) on 
$b_\bot(e)$, $b_\bot(N)$, $b_\bot(p)$, $b_\bot(\mu)$, $b_\bot(n)$ the couplings between a cosmic spin anisotropy 
field and different particles.}
\end{center}
\end{table}

The result Eq.~\eqref{resultA} can also be interpreted as a limit on the gravitational dipole moment of the neutron. 
In this case, $T$ is the solar period and the amplitude $A$ is expressed as
\begin{equation}
\label{A_gdm}
A = 2g_{\rm n} \frac{G M_{\odot}}{r_{\odot}^2} \frac{\cos(\lambda)}{h f_{\rm Hg}}, 
\end{equation}
where the inclination of the Earth with respect to the ecliptic plane, supressing the effect by less than 5\%, is neglected.
We thus obtain the following upper bound
\begin{equation}
|g_n| < 0.3 \ {\rm eV}/c^2 \, {\rm m} \quad  95 \% \ {\rm C.L.}
\label{ourgdm}
\end{equation}
In principle, much more stringent limits can be set using the Earth as a source of the new spin-dependent effect. 
In this case, one has to search for a steady signal contribution $\delta R$ added to or substracted from 
the main coupling term ${\gamma_{\rm n}}/{\gamma_{\rm Hg}}$, see Eq.~(\ref{Eq1}), depending on the direction of the
magnetic field.
Previous measurements with the same apparatus constrained the difference of the values $R_0$, 
the value of $R$ for $\partial B / \partial z=0$, for magnetic field pointing upwards and downwards to ~\cite{Baker2006,Baker2007}
\begin{equation}
|\delta R |=\frac{1}{2} \, |R_{0\uparrow} - R_{0\downarrow} | < 1.6 \times 10^{-6} \quad  95 \% \ {\rm C.L.}
\end{equation}
Using the Earth's mass ($M_{\oplus}$) and radius ($r_{\oplus}$) in Eq.~(\ref{Eq0}),
this can be translated into a limit on the neutron gravitational dipole moment
\begin{equation}
|g_{\rm n}| = |\delta R| \frac{h f_{\rm Hg}}{2} \frac{r_{\oplus}^2}{G M_{\oplus}} < 2.5 \times 10^{-4} \ {\rm eV}/c^2 \, {\rm m} \quad  95 \% \ {\rm C.L.}
\label{gdm}
\end{equation}

Finally, the limits derived above can be discussed in terms of the Hari-Dass framework of spin dependent gravity \cite{HariDass}:
\begin{equation}
V_{\mbox{Hari-Dass}} = \alpha_1 G M \frac{\hbar}{2c} \frac{{\bf \sigma} \cdot {\bf r}}{r^3} + \alpha_2 G M \frac{\hbar}{2c^2} \frac{{\bf \sigma} \cdot {\bf v}}{r^2} 
\end{equation}
where $\alpha_1$ and $\alpha_2$ are dimensionless parameters and $v$ is the neutron velocity with respect to the source.
$\alpha_1$ is directly proportional to the neutron gravitationnal dipole moment and the limit (\ref{gdm}) leads to:
\begin{equation}
|\alpha_1|=\frac{2 |g_n| c^2}{\hbar c} < 2.5\times 10^3 \quad  95 \% \ {\rm C.L.}
\end{equation}
this is the best limit for the neutron although still far above the natural value $\alpha_1 \approx 1$. 
A more stringent limit $|\alpha_1| \lesssim 2 \times 10^2$ was obtained using $^{199}$Hg and $^{201}$Hg~\cite{Ven92}, however involving nuclear model uncertainties.

While $\alpha_1$ is best constrained using the Earth as a source, the search for daily modulation is the natural way to probe $\alpha_2$, using the Sun as the source. 
Our limit (\ref{ourgdm}) translates to
\begin{equation}
|\alpha_2| < 3 \times 10^{10} \quad  95 \% \ {\rm C.L.}
\end{equation}
As for the limit on the cosmic anisotropy field, this is the first limit on the free neutron. 

We are grateful to the ILL staff for providing us
with excellent running conditions and in particular acknowledge the outstanding support of T.~Brenner. 
We also benefitted from the technical support throughout the collaboration.
We acknowledge fruitful discussions with R.~Lehnert.
This work is supported by the French-Polish funding agencies (French: convention IN2P3-Laboratoires Polonais, grant No. 05-120 and Polish:
COPIN/283/2006), the original development of the nEDM apparatus was funded by the UK's PPARC (now STFC) 
and the Cs magnetometers were financed by the Swiss National Science Foundation (200020--119820). 


\end{document}